\documentstyle[preprint,aps]{revtex}
\input psfig

\begin{document}
\sloppy
\title { Rotational alignment near N=Z and proton-neutron correlations \\ }
\author{S. Frauendorf$^{(1)}$ and J.A. Sheikh$^{(1,2,3)}$}
\address{$^{(1)}$IKHP,Research Center Rossendorf, PF 510119, 
D-01314 Dresden, Germany\\
$^{(2)}$Department of Physics, University of Surrey,
Guildford, Surrey GU2 5XH, UK\\
$^{(3)}$Tata Institute of Fundamental Research, Bombay,
400 005, India}

\maketitle

\begin{abstract}

The effects of the residual proton-neutron interactions 
on bandcrossing features are
studied by means of shell model calculations for nucleons in a
high-j intruder orbital. The presence of an odd-nucleon
shifts the frequency of the alignment of two nucleons of the other kind 
along the axis of rotation.  
It is shown that the anomalous delayed crossing observed in nuclei with
aligning neutrons and protons occupying the same intruder subshell can 
be partly attributed to these residual interactions.

\end{abstract}


\section{Introduction}

The cranked shell model (CSM) approach \cite{bf79,f80} accounts 
well for the overall
systematics of bandcrossing phenomena in rapidly rotating medium and
heavy mass nuclei. In this model the protons and neutrons move independently in
a fixed rotating potential. As a consequence of this assumption,
the alignment of a pair of neutrons with the rotational axis 
occurs at the same rotational frequency for different proton configurations.
However, recently substantial deviations from this simple picture
have been reported using the CSM
approach in the high-spin study of
nuclei near $N=Z$
 \cite{br73,br75,br77,kr72,kr74,kr75,kr76,kr79} and for nuclei 
with aligning neutrons and protons
occupying the same intruder subshell \cite{sb113,i117,cs119,re,wyss1,wyss2}.
These deviations, 
concerning the crossing between the g-band and the s-band in the 
even-neutron systems referred to as the (AB)-crossing,
 include:
\begin{itemize}
\item the crossing in the odd-proton nuclei is considerably 
delayed as compared to the neighboring even-even nuclei,
\item in the case of $N=Z$ even-even nuclei, the (AB)-crossing is
substantially delayed as compared to the neighboring $N\neq Z$.
\end{itemize}
Fig. \ref{f:crossex} summarizes the experimental evidence.

The observed delay of the neutron alignment 
has been attributed to an increase of the deformation induced by the odd
high-$j$ proton.  This mechanism has first been pointed out in ref.
\cite{f80} as the possible origin of
 the delayed $i_{13/2}$ neutron
backbend if an $h_{9/2}$ proton is present. It has been
substantiated by systematic calculations of the equilibrium 
deformations for bands with an odd-proton in the 
$h_{11/2}$, $h_{9/2}$
and $i_{13/2}$ orbitals (cf. e. g. \cite{wyss1,i117,re,wyss2}). Though
 being in the right order of magnitude, the calculations tend to
underestimate the experimental shifts.
The p-n residual interaction has been invoked to be responsible for 
the remaining discrepancy \cite{wyss1,i117}. In a 
study of a QQ- type 
p-n interaction acting in the particle-hole channel ref. \cite{wyss2} 
finds only very small shifts. In the pesent paper we further investigate the 
role of the residual interaction. At variance with ref. \cite{wyss2},
we use a short range force and calculate the exact solutions for the
model system of interacting protons and neutrons in a $j$ shell,
exposed to a deformed potential.

The strong short-range proton-neutron
(p-n) attraction
probes  the relative orientation of the high-$j$ orbitals, because of their
strongly anisotropic, torus like density distributions. 
It favors the anti parallel ($J=0,1$) and the parallel ($J=2j$)
coupling of a p-n pair, the energies being comparable. 
Since there are many 
low-$J$ pairs available, they tend to 
form a correlated state (p-n pairing). Processes in which the
high-$j$ particles change their relative orientation should be
good means to study these aspects of the p-n interaction. 
In this manuscript, we  investigate  a special kind of such reorientation,
the alignment of a pair of neutrons with the axis of rotation.  
By means of a shell model calculation it will be demonstrated that
the rotational frequency at which the neutron alignment occurs 
changes when additional protons are present and that the frequency shift
are sensitive to the p-n interaction and the correlations it generates.
It will also be shown that the  p-n interaction causes a delay
of the first double-bandcrossing in $N=Z$ model systems studied as compared to 
cases with  $N\not=Z$ which may be related to  the delayed bandcrossing
observed in $N=Z$ nuclei \cite{kr72}.


\section{The Shell Model}
In our model the protons and neutrons in a $j$-shell
interact via a delta force and move in the deformed 
rotating potential generated by the nucleons outside the $j$-shell.
The model hamiltonian is the same as used in refs.
\cite{she90,fsr94}, where one may look for the details, and is given by
\begin{equation}
 H^\prime = 
 - \sum_i \biggr[ 4({4\pi \over 5})^{1/2} \kappa Y_{20}(\hat r_i)
          +\omega j_i^{(x)}\biggr] - G \sum_{i<j} \delta(|\hat r_i-\hat r_j|).
\end{equation}
The first term is the deformed quadrupole field where
$\kappa$ is related to the 
deformation parameter $\beta$ by $\kappa = 51.5A^{1/3} \beta$ in 
units of the coupling strength $G$ of the delta force.
The second, cranking term describes the uniform rotation 
about the x-axis with the frequency $\omega$ and the third term is the
interaction.
The eigenfunctions of the hamiltonian are found by numerical diagonalization.
They can be classified by the isospin and the signature (symmetry
with respect to $R_x(\pi)$) which are conserved by the model hamiltonian.

We would like to
point that the present model predictions
should only be considered as qualitative indications of the physics expected
from more realistic studies.
The major restriction of our model is that
we consider only few particles occupying the intruder subshell with rest
of the nucleons giving rise to a deformed potential.
The assumption that the nucleons outside the intruder shell do not
take part in the correlations generated by the short range
residual interaction seems to be the most serious restriction of the model.  
Since we choose a realistic
value of the deformation for a given nucleus and estimate the number of
particles occupying the high-$j$ orbital by  looking at the realistic Nilsson 
diagrams, the estimates of the frequency where the rotational
alignment occurs should be qualitatively right.
As a consequence of the restriction to a single $j$-shell, 
the total angular momentum is underestimated. This is of little importance,
because we study the {\em frequency} of the alignment processes,
which will not be changed by the missing angular momentum from the nucleons
outside the $j$-shell.

\section{Examples for delayed crossings}

\begin{figure}[t]
\vspace*{-1cm}
\mbox{\psfig{file=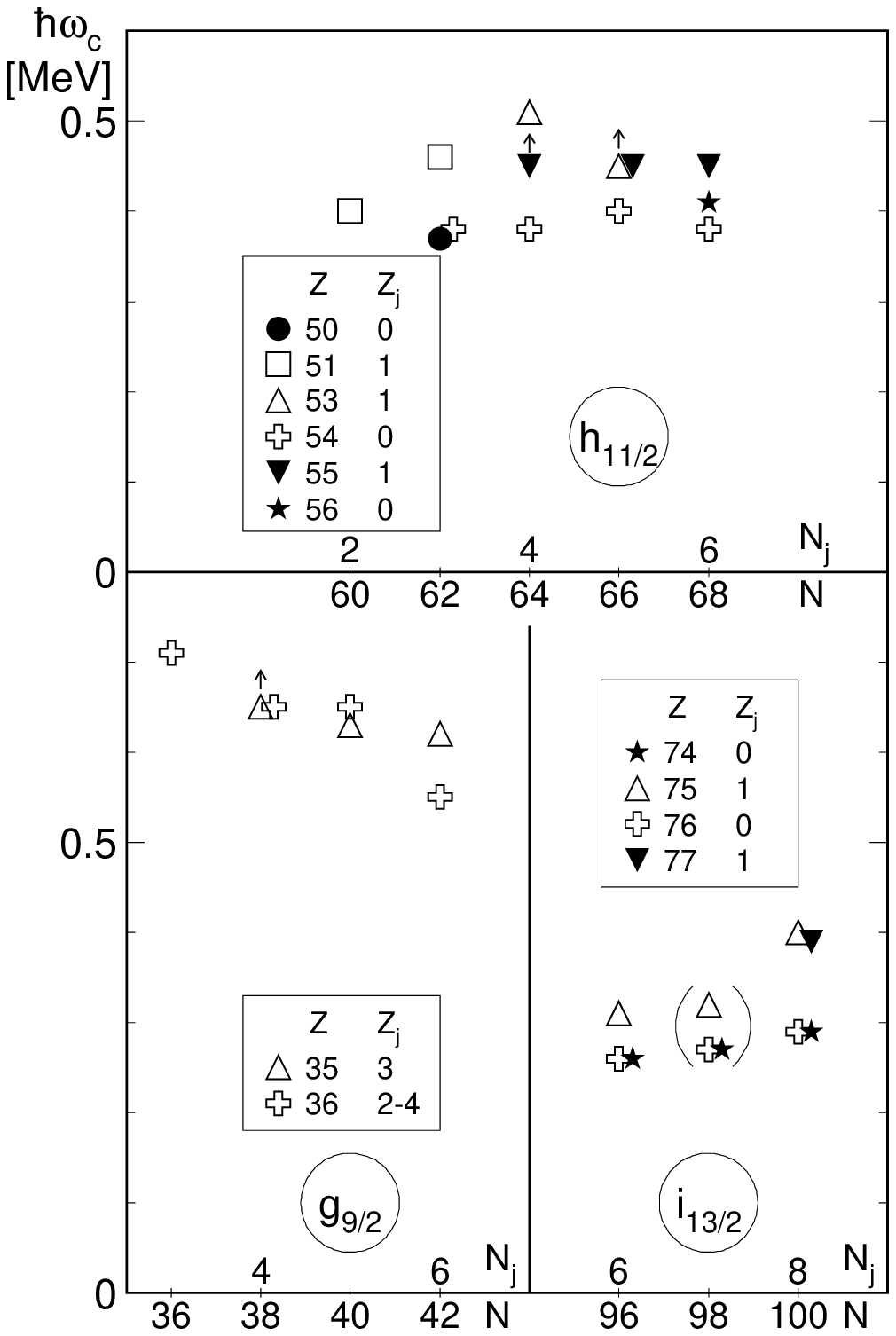,width=14cm}}
\vspace*{-1cm}
\caption{\label{f:crossex}
Experimental crossing frequencies, $\hbar \omega_c$, for the first
alignment of a pair of neutrons in the $g_{9/2},~h_{11/2}$ and $i_{13/2}$
shells. The odd-proton occupies  the same $j$  shell as the neutrons.
The number
$N_j$ is an estimate of the number of neutrons in the $j$  shell,
obtained by counting Nilsson levels at the deformation $\beta=0.25$.
For the $h_{11/2}$   and $i_{13/2}$  shells the proton occupancy
is zero or one, for the $g_{9/2}$ shell it is close to the neutron
occupancy.
The arrows indicate that only lower limits are known. For
$N=98$ the very gradual rise of the function $I(\omega)$ does not permit
a reliable determination of $\hbar \omega_c$, however the shift of the
two curves gives a reasonable estimate of the difference between the
crossing frequencies. In the case of $Z=50$ and 51 the deformed excited
excited bands are shown. The experimental data are from ref.s
\protect\cite{br73}-\protect\cite{re}.
The even Sr - isotopes and the $g_{9/2}$ bands of the Rb - 
isotopes do not show a $g_{9/2}$ alignment what might be a consequence of the
$Z=N=38$ gap in the single particle spectrum. 
}
\end{figure}

\begin{figure}[t]
\mbox{\psfig{file=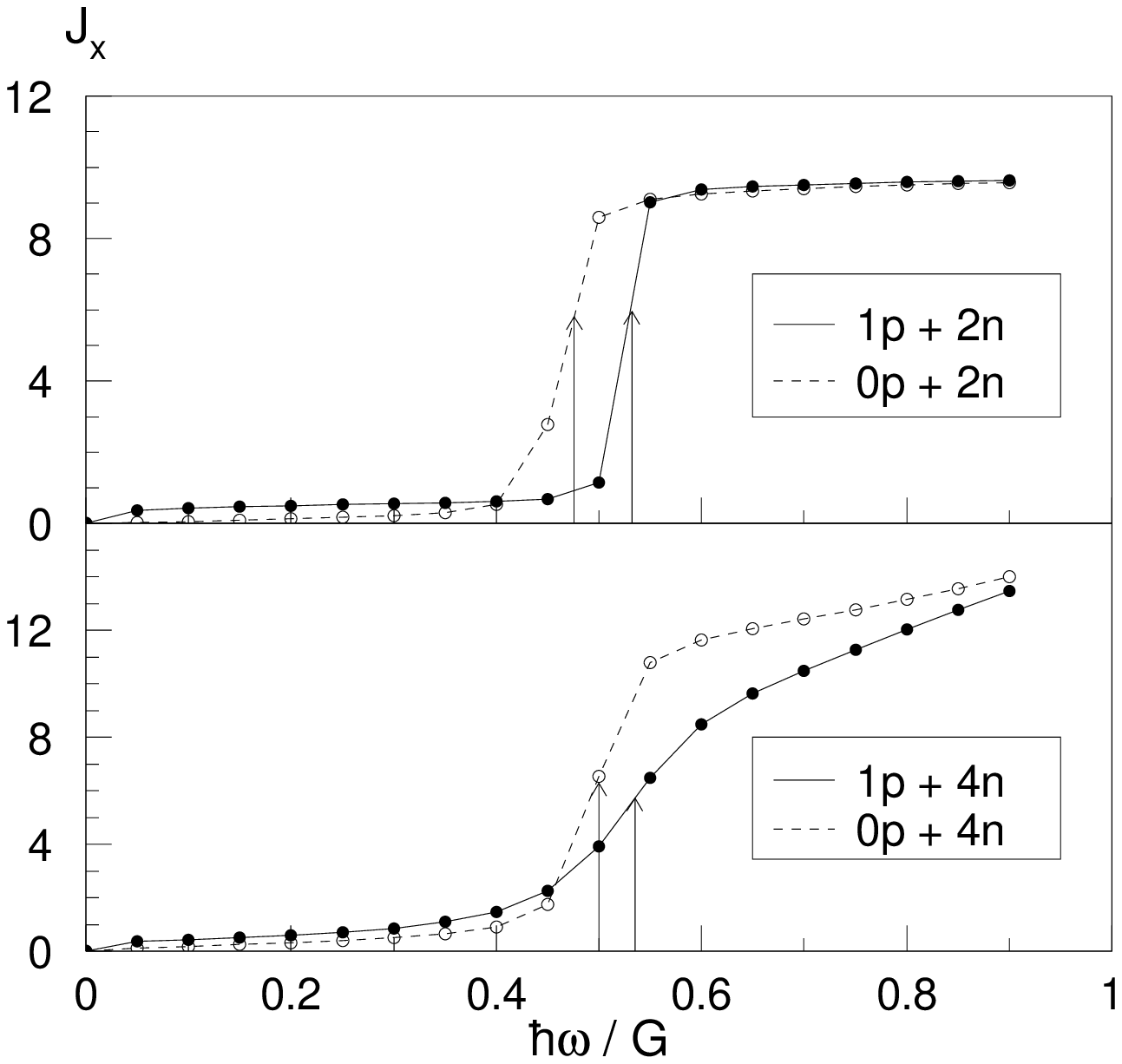,width=14cm}}
\caption{\label{f:crossdef11}
Angular momentum   $J_x$ as a function of the rotational frequency  for 
 particles in a $h_{11/2}$ shell with deformation $\beta$=0.25. The arrows
indicate the crossing frequencies. For odd-$Z$ a constant of 11/2 is subtracted
from $J_x$.}
\end{figure}

Fig. \ref{f:crossdef11} 
shows the expectation value of the $x$ component of the
angular momentum  calculated by assuming  
a deformation $\beta=0.25$ of a well deformed nucleus. The situation
with few protons and neutrons occupying the $h_{11/2}$ shell is studied.
The alignment of a 
pair of neutrons with the axis of rotation (x), caused by the
Coriolis force,
shows up as the step like rise of 
the angular momentum from small values to about 10.
It corresponds to the crossing of g- band with the neutron s-band, where
the inflection point defines the crossing frequency.
It is seen that the presence of the odd proton delays the alignment
of the neutron pair. As demonstrated by 
fig. \ref{f:crossex} both in the 
$h_{11/2}$ and in the $i_{13/2}$ shells a similar  delay of the crossing
frequency is seen
for the combination ($Z=1, N=2,4,6$).
The estimated number of protons and neutrons in the high-$j$ shell,
denoted by $Z_j$ and $N_j$, must be compared with 
$Z$ and $N$ of the calculations, respectively.

\begin{figure}[t]
\mbox{\psfig{file=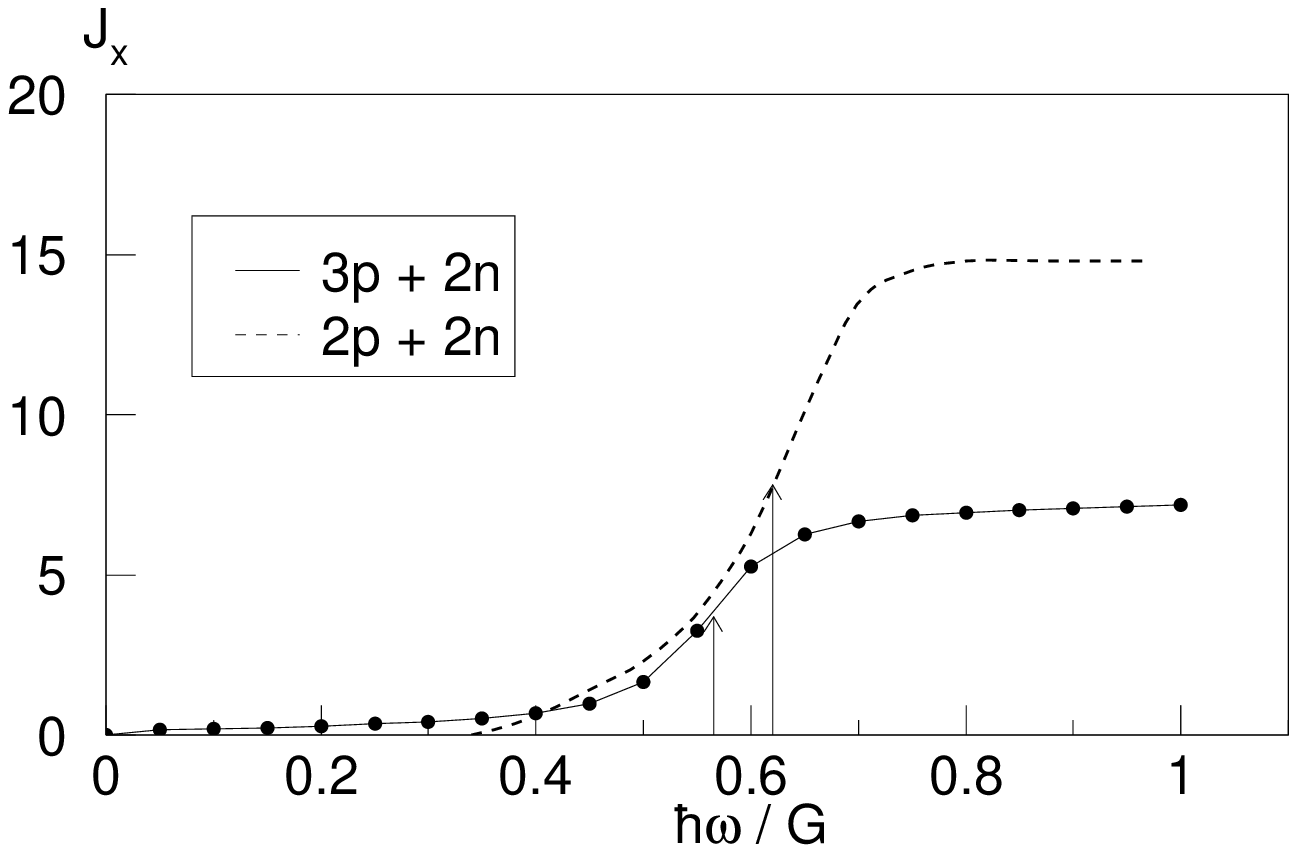,width=14cm}}
\caption{\label{f:crossdef9}
Angular momentum   $J_x$ as a function of the rotational frequency  for 
 particles in a $h_{9/2}$ shell with deformation $\beta$=0.25.
The case $Z=2, N=3$ is identical. The arrows
indicate the crossing frequencies. For odd-$Z$ a constant of 11/2 is subtracted
from $J_x$.}
\end{figure}

Fig. \ref{f:crossdef9} demonstrates that for near  symmetric filling 
of the $g_{9/2}$ shell ($Z=2, N=2$) and ($N=3, Z=2$)
the crossing frequency in the odd-A 
nuclei  lies below the first crossing in the 
even-even nuclei ($Z=N=2$), which corresponds to the
simultaneous alignment of protons and neutrons \cite{she90}.
This inversion of the order of the 
bandcrossings seems to be seen in the Br- and 
Kr-isotopes, as illustrated by  fig. \ref{f:crossex}.
The figure also shows that
the inverse ordering for near symmetric filling as compared to 
asymmetric filling can be attributed to a delay of the crossing
in the even even $N=Z$ nuclei
as compared to the ones with $N\not=Z$ \cite{kr72},  
which is stronger than the delay in the odd-$Z$ system.

\section{Systematics of the crossing frequency}
\begin{figure}[t]
\mbox{\psfig{file=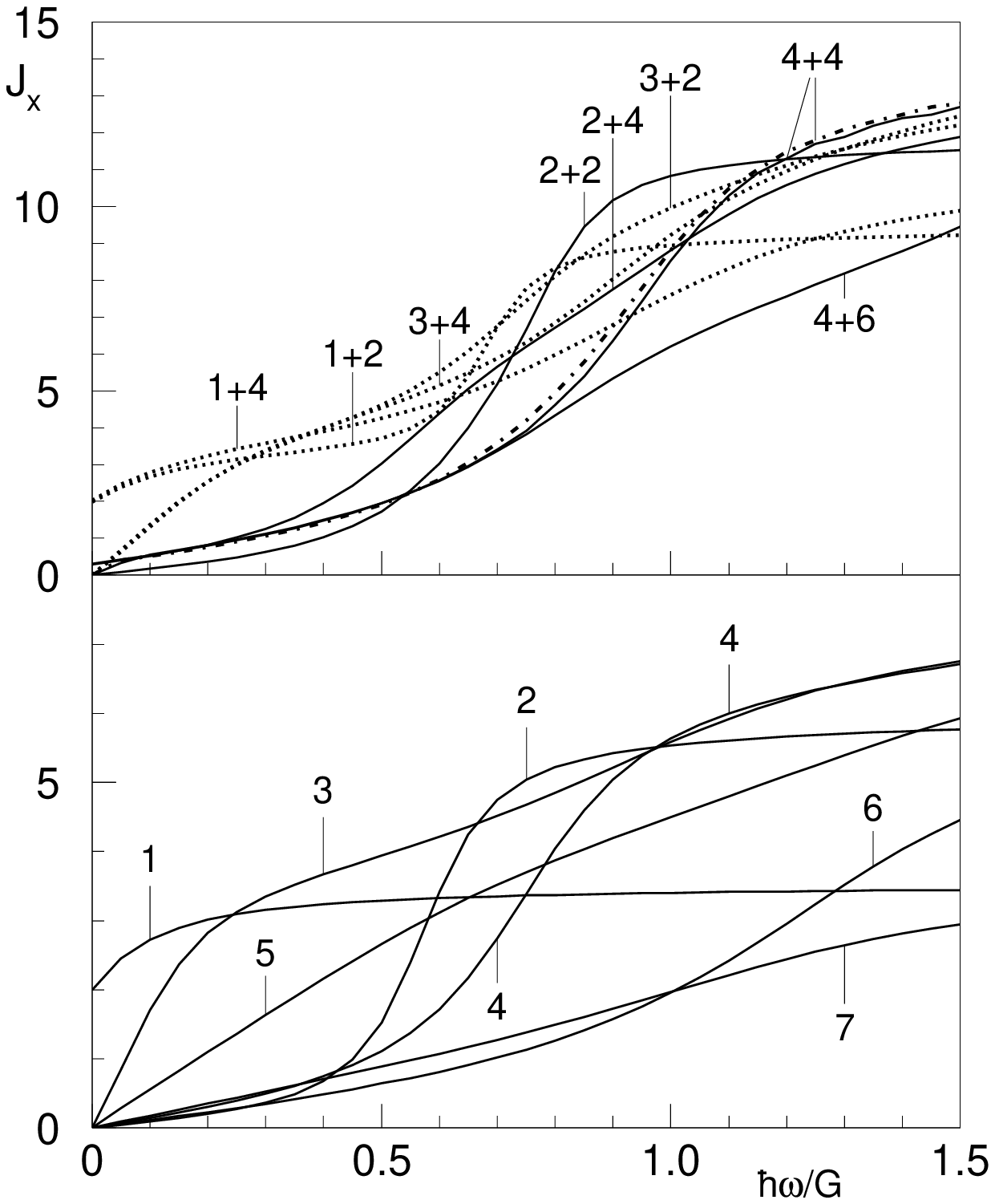,width=14cm}}
\caption{\label{f:f72jx} 
Angular momentum   $J_x$ as a function of the rotational frequency  for 
 particles in a $f_{7/2}$ shell with deformation $\beta$=0.25.
\\Upper panel: The particle numbers are quoted in the form $Z+N$.
Full lines: even $A$, dotted: odd-$A$, 
dashed dotted: only $T=1$ part of the $\delta$ force.
\\Lower panel: Only one kind of particles, the number of which is quoted. }
\end{figure}

\begin{figure}[t]
\mbox{\psfig{file=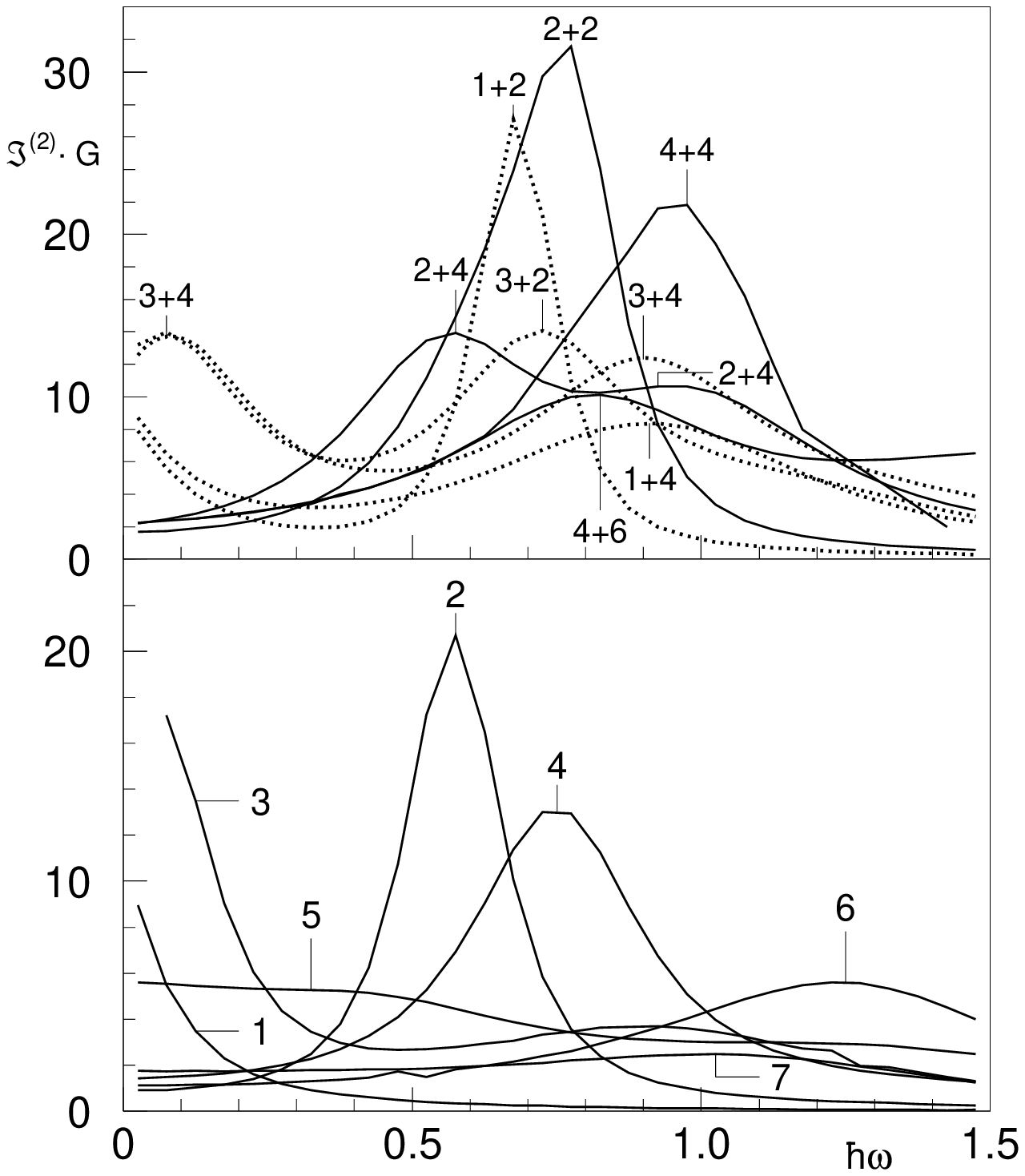,width=14cm}}
\caption{\label{f:f72momi} 
Dynamic moment of inertia   ${\cal J}^{(2)}$ 
as a function of the rotational frequency  for 
 particles in a $f_{7/2}$ shell with deformation $\beta$=0.25.
\\Upper panel: The particle numbers are quoted in the form $Z+N$.
Full lines: even $A$, dotted: odd-$A$. 
\\Lower panel: Only one kind of particles, the number of which is quoted. }
\end{figure}

The features discussed in the previous section reflect a systematic 
tendency that is
illustrated in fig. \ref{f:f72jx} and \ref{f:f72momi}, showing
$J_x(\omega)$ and ${\cal J}^{(2)}(\omega)=dJ_x/d\omega$ for the $f_{7/2}$
shell.  (We have chosen $f_{7/2}$ shell since the dimensions of the matrices 
to be diagonalized are lower and a systematic study is possible.) 
The alignments show up as peaks in ${\cal J}^{(2)}$.
We are interested in the alignment of a pair of $f_{7/2}$
neutrons, which is displayed in lower panels of the figs. for the pure neutron
$N=2$ and $N=4$ cases. If there were no p-n correlations,
the frequency at which these alignments appear would not change with the
number of protons. However, this is not seen. Instead, the critical 
frequency for
the neutron alignment first grows with the number of added protons.  
The shift reaches its maximum when $Z=N$ and then decreases
when $Z$ becomes larger than $N$. Of course, the role of $N$ and $Z$ 
could  be exchanged.

Qualitatively, one may ascribe the progressive delay of the crossing to 
p-n correlations that disfavor the simultaneous alignment of 
protons and neutrons (e. g. the presence of 
p-n pairs with a low $J$), generated
by the attractive p-n interaction. The more protons are added the stronger
the correlations become, until $Z=N$. 
As discussed in \cite{fsr94}, the character of the p-n interaction
between the aligned protons and neutrons
changes from attractive to repulsive when $Z$ exceeds $N$, because
the character of the aligned configuration changes from particle-like
to hole-like. This shifts the p-n correlations to the 
highly excited states, whereas the yrast states are not much modified
by the p-n interaction.   
  
The shift of the crossing frequency is a consequence of the $T=1$ part of
the $\delta$-interaction. The dashed dotted line in
fig. \ref{f:f72jx} shows a calculation where the $T=0$ part of the 
$\delta$-interaction has been removed. The function $J_x(\omega)$ almost
coincides with the one calculated with the full force. 
The question if the p-n correlations that cause the shifts
belong to the particle-hole channel or the particle-particle channel
(or to both)  cannot be decided on the basis of the shell model calculations.
For this, comparisons with mean field calculations are necessary, which
will be  addressed in a forthcoming paper.
We state here only the existence of strong p-n correlations
of the $T=1$ type between nucleons in the same $j$-shell, which cause
a delay  of the first bandcrossing. 

A delay of the onset of the bandcrossing for $N=Z=36$ has recently been
observed in the Kr isotopes \cite{kr72} (cf. fig. \ref{f:crossex}). 
It is suggested in the paper that
the delay may be a consequence of  $T=0$ pair correlations. 
As already mentioned, the dashed dotted line in fig. \ref{f:f72jx}
shows a calculation where we have switched off the
$T=0$ part of the interaction. It is seen that the results are almost the
same as for the full force, i.e. the $T=0$ correlations do not influence
the alignment significantly in the present model analysis and it is
interesting to see whether this conclusion will hold in a realistic
study.

Comparing in figs. \ref{f:f72jx} and \ref{f:f72momi}
the alignment of the odd-proton in the $Z=3$ system with the one in the 
$Z=3,~ N=2,~4$ one notices also the the alignment of an odd-proton is
hindered by the presence of neutrons in the same $j$-shell.  

In the
previous work \cite{she90} it was mentioned that strong p-n correlations
can alter the bandcrossing substantially and the double-bandcrossing
can occur below the individual proton and neutron crossings. However,
most of the analysis was performed in the spherical approximation and
the dominant component
of the p-n interaction which changes the bandcrossing was not fully explored. It
is shown in the present work using the proper deformed basis that 
it is $T=1$ component of the
n-p interaction which is responsible for the particle number dependence
of the bandcrossing.
This is in accordance with a recent mean field study \cite{kan} 
of the $j$-shell model system, which attributes the delay
of the crossing frequency in 
the even - even  $N=Z$ system  to the  $T=1$
monopole pair field. 

\section{Conclusions}
It is clear from the present simple model study that the alignment  of 
high-$j$ nucleons with the rotational axis is sensitive to
the p-n interaction. The alignment of one kind of particles
is delayed when the other kind of particles is present in the same $j$-shell.
The effect reaches its maximum for $N=Z$ 
and is a  consequence of   $T=1$  correlations
between protons and neutrons.
For $Z=N$ the crossing frequency in the even system is higher
than for asymmetric filling.
The experimentally observed delay of the first backbend in $Z=N$ even-even
$^{72}$Kr can be attributed to these correlations.
The experimentally observed
delay of the first backbend in even-neutron nuclei 
by the presence of an odd-proton in the same $j$-shell
could be a combination of these correlations and
the change of deformation induced by the odd-nucleon. By measuring the
deformation and/or calculating it one may try to disentangle
the two effects and probe the p-n  correlations. 
Finally, we would like to mention that while the results of the present
model analysis seem to point into the right direction they should
not be compared with the experimental data in a quantitative form. 
It would be quite
interesting to study the effects in
more realistic shell model configuration spaces.



\newpage


\end{document}